\documentstyle[epsfig]{article}

\newcounter{eqnletter}[equation]
\catcode`\@=11
\@addtoreset{equation}{section}   % Makes \chapter reset 'equation' counter.
\catcode`\@=12

\begin{document}

\begin{center}

{\LARGE\bf On the convergence of the usual \\[4mm]
perturbative expansions}

\vskip 1cm

{\large {\bf G.M. Cicuta} }

\vskip 0.1 cm

Dipartimento di Fisica, Universita di Parma,\\
and INFN Gruppo collegato di Parma,\\
Viale delle Scienze, I - 43100 Parma, Italy \footnote{E-mail
address: cicuta@parma.infn.it} \\[0.5cm]

\end{center}
\vspace{1 cm}
\begin{abstract}
The study of the convergence of power series expansions of energy
eigenvalues for anharmonic oscillators in quantum mechanics differs from
general understanding, in the case of quasi-exactly solvable potentials.
They provide examples of expansions with finite radius and suggest
techniques useful to analyze more generic potentials.
\end{abstract}

\vspace{1 cm}
{\bf Key words:} Perturbative expansions, anharmonic oscillators,
quasi-exactly solvable potentials, Bender-Dunne polynomials.
\vspace{1 cm}

\section{Introduction}
Let me summarize some of the most relevant features of perturbative
expansions for eigenvalues in one-dimensional quantum mechanics. Let us
consider the Schr\"odinger equation
\begin{eqnarray}
\left[ -\frac{\partial^2}{\partial x^2}  +\frac{x^2}{4}+
\lambda V(x)\right] \psi_k(x) =E_k \, \psi_k(x) 
\label{a.1}
\end{eqnarray}
where the harmonic oscillator is perturbed by a higher order even
polynomial $ \lambda V(x) =  \lambda p(x^2) $ and $k$ is the number of
zeros of the wave function .\\
In well known papers C.Bender and T.T. Wu \cite{bender1} and B.Simon
 \cite{simon} studied the quartic anharmonic oscillator
$p(x^2)= x^4$ .  The former authors \cite{bender1}
evaluated a large number of terms of the perturbative series
\begin{eqnarray}
E_k (\lambda)
= (k+ 1/2) + \sum_{n=1} ^\infty \lambda ^n E_k ^{(n)}
\label{a.2}
\end{eqnarray}
of the lowest energy eigenvalues $E_k (\lambda)$, $k=0,1,..$ and discussed the
occurrence of singularities in the complex 
$\lambda$ plane. It was found that for any energy level $E_k(\lambda)$
 there exist infinitely many points $\bar{\lambda}$
, which are  extrema of square root
branch singularities , the extrema accumulate
at the origin in the complex $ \lambda$ plane  and each value
$\bar{\lambda}$
 correspond to the crossing between pairs of energy levels
\begin{eqnarray}
E_n(\bar{\lambda})=E_{m}(\bar{\lambda})
\label{a.3}
\end{eqnarray} 
These features prevent a non vanishing radius of convergence for the
perturbative expansion of any energy eigenvalue (\ref{a.2}). The second
author \cite{simon} confirmed these results, by using Hilbert space
methods. Analogous occurrence of infinitely many singularities
with accumulation point at the origin, for any energy level, was found
if the quartic monomial $\lambda \, x^4$ was replaced by a higher
order monomial $\lambda \, x^{2n}$, $n>2$ \cite{simon}. T.Banks and
C.Bender \cite{bb} also studied the anharmonic oscillator with a
general polynomial potential (of even parity)
\begin{eqnarray}
H=-\frac{\partial^2}{\partial x^2}  +\frac{x^2}{4}+
g \left[ \left( \frac{x^2}{2} \right)^N +a \left(\frac{x^2}{2}
\right)^{N-1} +b \left(\frac{x^2}{2}\right)^{N-2}+... \right]
\label{a.4}
\end{eqnarray}
The $k-th$ energy level has a perturbative expansion
\begin{eqnarray}
E_k (g)= k+ \frac{1}{2} +\sum_{n=1}^\infty c_n ^{(k)} (a, b,..)g^n
\nonumber
\end{eqnarray}
They found that the large order behaviour of the coefficients
$ c_n ^{(k)} (a, b,..)$ is given by
\begin{eqnarray}
\frac{c_n ^{(k)} (a, b,..)}{c_n ^{(k)} (0, 0,..)} \sim e^{a/(N-1)}
\left[ 1+ O(\frac{1}{n}) \right]
\label{a.5}
\end{eqnarray}
Then the singularities of $E_k (g)$ closest to the origin in the complex
$g$ plane are controlled by the highest order monomial $g (x^2 /2)^N$
whereas the next order monomial $g a (x^2 /2)^{N-1}$ only results in a 
constant factor and the next order monomial $g b (x^2 /2)^{N-2}$
affects the corrections of order $O(1/n)$ with respect to the previous
result.\\
For some decades it was believed that any formal Taylor expansion
of energy eigenvalues of an anharmonic potential, with any polynomial
perturbation (of degree higher than quadratic) would have a vanishing 
radius of convergence.
It was recently found that for the class of models
known as quasi exactly solvable potentials, a number of energy levels
have a perturbative expansion with finite radius of convergence
\cite{book}.
The singularities of these energy levels still correspond to level
crossing, yet these are a finite number. Quantum mechanics being
a $(0+1)$ dimensional quantum field theory, it would be exciting to
find similar convergence in higher dimensional models of
quantum field theory.\\
Further references to extensive investigations on the divergence of
 the perturbative expansion in quantum mechanics and in quantum field
theory may be found in \cite{arteca} and \cite{zinn}.\\
It is clear that quasi-exactly solvable models have convergent
perturbative expansions for a finite number of energy eigenvalues
because those eigenvalues are decoupled from the rest of the spectrum.
Yet this property is so peculiar, that it is interesting to have a
pattern of the radius of convergence as function of the parameters.
This is evaluated in sect.2, in an algebraic exact fashion, for a
simple sequence of potentials. It also seems that quasi-exactly solvable
 potentials provide efficient tools to investigate the possibility
of convergence for generic potentials, when these are summed in a
fashion similar to quasi-exactly solvable models. This analysis is presented
but not completed in sect.3.

\section{Quasi-exactly solvable potentials.} 
The generic conclusion of divergence of perturbative expansion does not
hold in the case of quasi-exactly solvable potentials. This is
clearly stated in the book \cite{book}. In this section it will be
exhibited by the evaluation of the radius of convergence in a sequence
of cases.\\
The simplest class of quasi-exactly solvable potentials corresponds to 
the one-dimensional quantum sextic oscillator model with hamiltonian
\begin{eqnarray}
H_M=-\frac{\partial^2}{\partial x^2}+[b^2-a(4M+3)]x^2+2abx^4+a^2x^6
\label{a.11}
\end{eqnarray}
where $a$ is  positive, $b$ is real, $M$ is a non-negative
 integer ($M=0, 1, ..)$
For sake of a simpler exposition, let us choose $b=1$ (which is
a generic value). It can be shown that the eigenvalue equation
\begin{eqnarray}
\left[-\frac{\partial^2}{\partial x^2}+[1-a(4M+3)]x^2+2ax^4+a^2x^6]
\right] \psi_k (x)=E_k \psi_k (x)
\label{a.12}
\end{eqnarray} 
where $\psi_k (x)$ is square integrable, 
has the lowest part of the spectrum corresponding to the
even wave functions, which may be computed in closed form in
algebraic way. That is, the first $M+1$ even wave functions
$\psi_{2k} (x^2)$, $k=0, 1, ..M$ are
\begin{eqnarray}
\psi_{2k} (x^2) &=& e^{- \frac{x^2}{2}-\frac{ax^4}{4}}
\prod_{i=1}^M \left( \frac{x^2}{2}-\xi_i \right)=
\label{a.13}\\
&=& e^{- \frac{x^2}{2}-\frac{ax^4}{4}}
\sum_{n=0} ^M  \frac{ (-1)^n P_n(E)}{(2n) !}x^{2n}
\label{b.13}
\end{eqnarray}
where $\xi_1,\dots, \xi_M$ are real numbers satisfying
the system of $M$ algebraic equations
\begin{eqnarray}
\sum_{k=1,k\neq i}^M\frac{1}{\xi_i-\xi_k}+\frac{1}{4\xi_i}-1-2a\xi_i=0
\label{a.14}
\end{eqnarray}
Each of the $M+1$ solutions of the set $\{\xi_i\}$ is characterized by
having $k$ of the numbers  $\{\xi_i\}$ positive and the remaining $M-k$ being
negative. It provides one of the $M+1$ computable energy eigenvalues:
\begin{eqnarray}
E_{2k}=(4M+1)+8a\sum_{i=1}^M \xi_i
\label{a.15}
\end{eqnarray}
and the ground state corresponds to the solution where all the numbers
 $\{\xi_i\}$ are negative. The eq.(\ref{a.14}) has the familiar form of a
saddle point equation for random matrix models and the promising relations
between quasi exactly solvable models and random matrix models are just 
beginning to be explored \cite{cu} \cite{csu}.\\
The system (\ref{a.14}) and eq.(\ref{a.15}) imply that for fixed non negative
integer $M$, the  $M+1$ eigenvalues of the lowest even wave functions
are the roots of a polynomial equation of order $M+1$ which may be obtained
by techniques of symmetric functions. It is however easier to obtain
them by inserting the ansatz (\ref{a.13}) in the eigenvalue equation 
(\ref{a.12}) bypassing the evaluation of the set  $\{\xi_i\}$.
For example, the five polynomial equations which correspond
to the values $M=0, 1,..,4$ are
\begin{eqnarray}
& & E-1=0 \nonumber \\
& & E^2-6 E+5-8 a=0 \nonumber \\
& & E^3-15 E^2+(59-64a)E+(192a-45)=0   \nonumber \\
& & E^4-28 E^3+(254-240a)E^2-(812-2592a)E+(585-4656a+2880 a^2)=0  \nonumber \\
& & E^5-45 E^4+10(73-64a)E^3-114(45-128a)E^2+ \nonumber \\
& & \qquad 128(\frac {14389}{128}-
   709a+368 a^2)E-128 (\frac {9945}{128}-984a+1776 a^2)=0 
\label{a.16}
\end{eqnarray}
The second ansatz for the wave function (\ref{b.13}) is very useful
to derive the wave function because one immediately finds a three term
recursion relation for the coefficients $P_n(E)$
\begin{eqnarray}
(E-4k-1)P_k=P_{k+1}-8ak(2k-1)(k-1-M)P_{k-1}
\label{a.17}
\end{eqnarray}
with
\begin{eqnarray}
P_0=1 \; , \qquad P_1=E-1
\label{a.18}
\end{eqnarray}
The coefficient $P_n(E)$ is then a  polynomial in $E$ of order $n$.
The condition that $P_{M+1}(E)=0$ leads to the algebraic equation for $E$
of degree $M+1$ (the lowest ones being eq.(\ref{a.16}) ). This
condition and the recursion relation eq.(\ref{a.17}) imply that
all $P_k(E)$ with $k>M$ vanish. The finite set of non-vanishing
 polynomials $P_n(E)$ 
is a set of weakly-orthogonal polynomials, recently discussed
by several authors \cite{bd}, \cite{alex}, \cite{finkel}. 
The papers \cite{alex}, \cite{finkel} show that 
a set of weakly-orthogonal polynomials occur in any
quasi-exactly solvable model and they exhibit
 the discrete weight function 
$ \omega (E)=\sum_{k=0} ^M \omega_k \delta (E-E_k)$
for which
\begin{eqnarray}
\int { P_n (E) P_m (E) \omega (E) dE} = h_n \delta_{n, m}
\label{a.19}
\end{eqnarray}
The roots of the polynomial eqs.(\ref{a.16}), $P_{M+1}=0$, $M=0, 1,...$, 
determine the even lowest energy eigenvalues $E_0 (a), \, E_2(a), \,..,
E_{2M} (a)$. The singularities of the function $E_{2k} (a)$ closest to
the origin, in the complex plane of the variable $a$ provide the radius
of convergence of the perturbative expansions
\begin{eqnarray}
E_{2k} (a)= 1 + 4 k + \sum_{n=1} ^\infty d_n ^{(k)} a^n
\label{a.20}
\end{eqnarray}
The singularities only occur for the values $ {\overline a}$ such that
$ E_{2k} ({\overline a})$ is a multiple root and may be found by
examining the solution of the system
\begin{eqnarray}
P_{M+1}(a,E)=0=\frac {\partial}{\partial E}P_{M+1}(a,E)
\label{a.21}
\end{eqnarray}
For instance, for $M=2$,
\begin{eqnarray}
P_3(E)= E^3-15 E^2+(59-64a)E+(192a-45)=0
\label{a.22}
\end{eqnarray}
defines, in closed form, the three energy levels
$E_0 (a), \, E_2 (a), \, E_4 (a)$, and their perturbative expansions
may be easily evaluated at arbitrary order
\begin{eqnarray}
E_0 (a)=1-4a-2a^2+4a^3+\frac{1}{2}
 a^4-11a^5+\frac{39}{4} a^6+\frac{57}{2} a^7-\frac{2235}{32} a^8-
\nonumber \\
\frac{563}{16} a^9+\frac{21809}{64} a^{10}+...
\label{a.221}
\end{eqnarray}
\begin{eqnarray}
E_2 (a)=5-8a+32a^2-160a^3+1024a^4-7552 a^5+59904 a^6-497664 a^7+
\nonumber \\
+4276224 a^8- 37697536 a^9+339042304 a^{10}+...
\label{a.222}
\end{eqnarray}
\begin{eqnarray}
E_4 (a)=9+12 a-30a^2+156 a^3-\frac{2049}{2} a^4+7563 a^5-\frac{239655}{4}a^6+
\nonumber \\
+\frac{995271}{2} a^7-\frac{136836933}{32}a^8+\frac{603161139}{16}a^9-
\frac{21698729265}{64}a^{10}+...
\label{a.223}
\end{eqnarray}
The singularities of $E_0 (a)$ , eq.(\ref{a.22}), occur for the three values
${\overline a}$ 
\begin{eqnarray}
{\overline a} &=& [3(11+64 \sqrt{3})^{1/3} -7- 
\frac{69}{(11+64 \sqrt{3})^{1/3}}]/64
\sim -0.09445127  \nonumber \\
{\overline a} &\sim & -0.116837 \pm 0.389587 \, i
\label{a.23}
\end{eqnarray}
where the above pair of complex conjugate values are the roots of
\begin{eqnarray}
{\overline a}^2 &+& \frac{7406+1587 \rho +(33 -192 \sqrt{3})\rho ^2}{33856} 
{\overline a}+
\frac{\rho}{69+7 \rho-3 \rho ^2}=0
 \nonumber \\
 \rho &\equiv& (11+64 \sqrt{3})^{1/3}
\label{a.231}
\end{eqnarray}
It is easy to check that the closest, real negative value of ${\overline a}$ 
, (\ref{a.23}), corresponds to the radius of convergence of the perturbative
expansions of both the levels $E_2 (a)$
and $E_4 (a)$, (\ref{a.222}) and (\ref{a.223}),
and may be interpreted as the value of $a$ corresponding to
the crossing $E_2 ({\overline a})=E_4 ({\overline a})$
, while the couple of complex conjugate values (\ref{a.231})
correspond to the radius of convergence of the 
perturbative expansion of the ground level $E_0 (a)$ , (\ref{a.221}).
It is remarkable how easily this situation generalizes for all 
integer values of $M$. The $M+1$ roots  of the polynomial equation
$P_M(E)=0$ define the $M+1$ energy eigenvalues $E_0(a), \,E_2(a), \,...
E_{2M}(a)$;  the system (\ref{a.21}) leads to a polynomial equation
in the variable $a$ of degree $ M(M+1)/2$. Its $ M(M+1)/2$ roots in the
 complex $a$ plane are in one-to-one correspondence with the possible
level crossing among pairs of eigenvalues
\begin{eqnarray}
E_{2r}({\overline a})=E_{2s}({\overline a})
\label{a.24}
\end{eqnarray}
All these singular values were examined, for the cases $M=1$ up to $M=7$,
 beginning with the values ${\overline a}$ with the smallest modulus.
For any $M$ examined, the first value ${\overline a}$ occurs on the
real negative axis in the $a$ complex plane, it corresponds to
crossing of the two highest eigenvalues considered 
(for $M=7$ it is $E_{12}({\overline a})= E_{14}({\overline a})$)
 thus providing the radius of
convergence of their perturbative expansions (\ref{a.20}).
Singular values ${\overline a}$ with larger modulus describe  
level crossing between pairs of intermediate levels. Only the
$M$ values  ${\overline a}$ with larger modulus describe
level crossings of the ground state level with the other $M$ levels.
The  radius 
of convergence of the perturbative expansion of the ground level $E_0 (a)$
is determined by a couple of  complex conjugate values of ${\overline a}$
with the smallest modulus in this last group of $M$ values ${\overline a}$.
This analysis is confirmed by the study of the period of oscillations
of the coefficients in the perturbative expansion of $E_0 (a)$.\\

\begin{figure}[t]
\begin{center}
\mbox{\epsfig{file=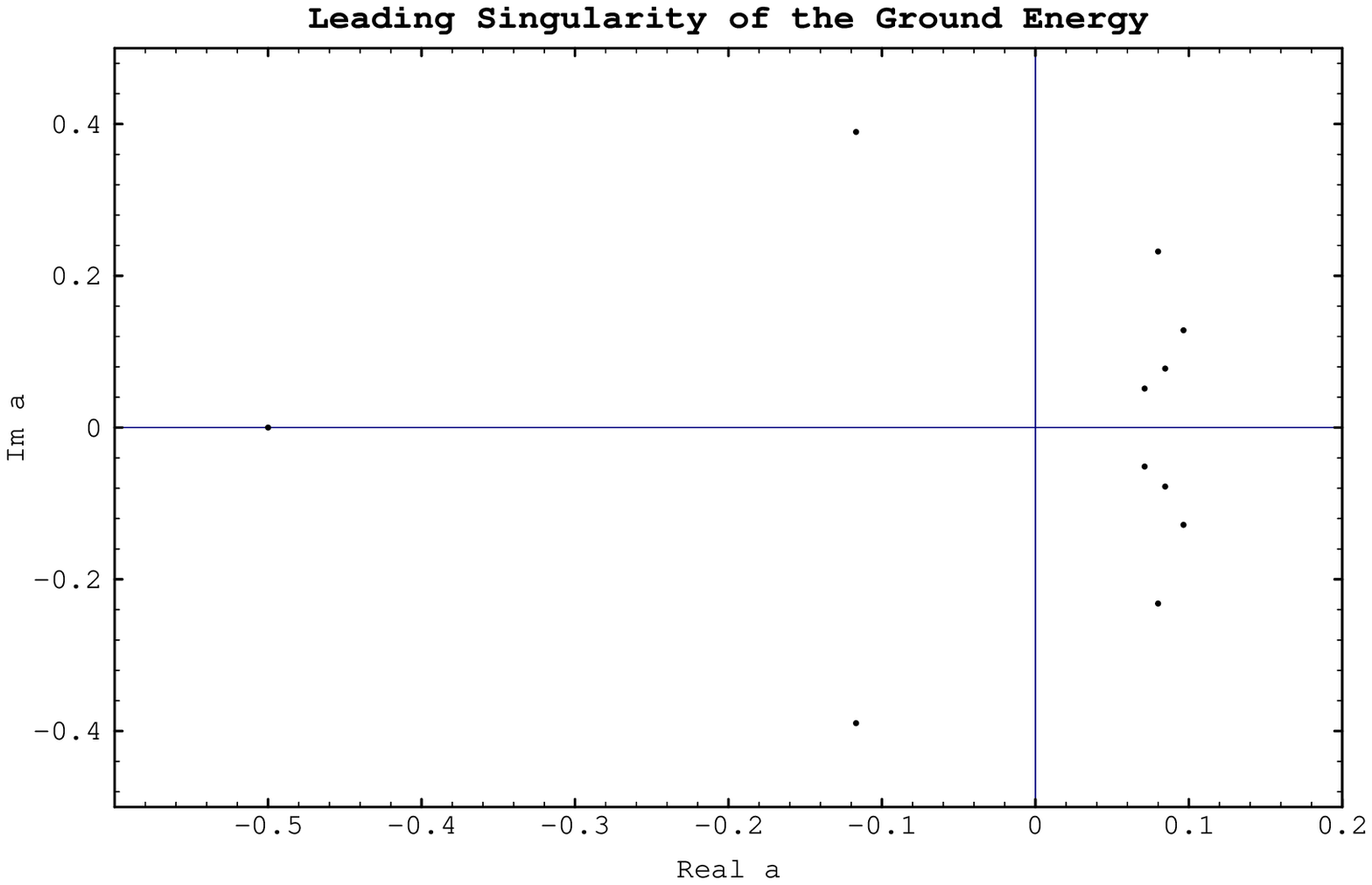,height=6.5cm}}
\end{center}
\caption{}
\end{figure}

Fig.1 shows a sequence of points ${\overline a}$ in the complex $a$ plane.
The point ${\overline a}=-1/2$ corresponds to the only singularity of the
ground state eigenvalue $E_0 (a)=3-2 \sqrt {1+2a}$ of the potential
(\ref{a.12}) with $M=1$.
 Moving clockwise in the upper half plane, there are points 
corresponding to the  singularity of $E_0 (a, M)$ closest to the origin,
for the cases $M=2, 3, ..,6$. One sees that the radius $r_M$ of convergence
decreases as $M$ increases.

\section{Softly broken quasi-exact solvability}

The analysis of quasi-exactly solvable models is useful for more general
polynomial potentials, where the couplings do not obey the constraints
of quasi-exact solvability.
Let us  consider the hamiltonian with a generic sextic potential

\begin{eqnarray}
H=-\frac{\partial^2}{\partial x^2} + {\cal V}(x)=
-\frac{\partial^2}{\partial x^2} +\alpha x^2 +
\beta x^4 +\gamma x^6
\label{c.1}
\end{eqnarray}
with $\gamma >0, \, \alpha$ and $\beta$ real. By choosing
\begin{eqnarray}
a=\sqrt{\gamma}  \, , \;b=\frac{\beta}{2 \sqrt{\gamma}}\, , \;
M=\frac{1}{4}[\frac{1}{\sqrt{\gamma}}(\frac{\beta ^2}{4 \gamma}-\alpha)-3]
\label{c.2}
\end{eqnarray}
the hamiltonian (\ref{c.1}) reproduces the quasi-exactly solvable
model (\ref{a.11}), but now $M$ is real, rather than a non negative
integer. In this Section, I study the perturbation theory for 
the eigenvalues of this hamiltonian, by keeping $b$ and $M$ fixed,
in series of powers of the coupling $a$. This is just a specific
way to do perturbation theory for the generic sextic potential
(\ref{c.1}). For simplicity let us fix again $b=1$, which is a 
value with no special meaning, 
and let us choose the (formal) ansatz for the even parity
wave function 
\begin{eqnarray}
\psi (x^2) &=& \phi (a,x^2) \;
e^{- \frac{x^2}{2}-\frac{ax^4}{4}} \nonumber \\
\phi (a,M,x^2)&=&
\sum_{n=0} ^{\infty}  \frac{ (-1)^n P_n(E,a,M)}{(2n) !}x^{2n}
%E=1+ \sum_{k=1} ^{\infty} c_k a^k
\label{c.3}
\end{eqnarray}
The eigenvalue eq.(\ref{a.1}) implies the recursion relations
\begin{eqnarray}
P_n(E, a, M)=(E-4n+3)P_{n-1}(E, a, M)-\lambda_n P_{n-2}(E, a, M)
\nonumber \\
\lambda_n \equiv 8a(n-1)(2n-3)(M+2-n)
\label{d.1}
\end{eqnarray}
Since $M$ is not a non-negative integer, the number of non-vanishing
polynomials  $P_n(E,a,M)$ is infinite. However $\lambda_{n+1}>0$
only for  $n<M+1$ so the system of infinite polynomials  $P_n(E,a,M)$
is not an orthogonal systems with respect to a non-negative
Stieltjes measure \cite{chihara}. Still the recursion relations
(\ref{d.1}), are a powerful tool for the perturbative analysis
of the generic sextic potential (\ref{c.1}).
The first few polynomials $P_n(E,a,M)$ are
\begin{eqnarray*}
P_0(E,a,M)= 1 \; ; \;
P_1(E,a,M)= E-1 
\end{eqnarray*}
\begin{eqnarray*}
P_2(E,a,M)= E^2-6E+5-8aM 
\end{eqnarray*}
\begin{eqnarray*}
P_3(E,a,M)= E^3-15E^2+(59-56aM+48a)E+(120a M-48a-45)  
\end{eqnarray*}
\begin{eqnarray*}
P_4(E,a,M)\hspace{-1em}&=\hspace{-1em}& E^4-28E^3+(254+288a-176a M)E^2-
(812+2112a-1568a M)E+ \\
 \hspace{-1em}&+\hspace{-1em}& (585+1824a-2160a M-1920 a^2 M+960 a^2 M^2)
\end{eqnarray*}
\begin{eqnarray}
P_5(E,a,M)\hspace{-1em}&=\hspace{-1em}& E^5-45E^4+10(73+96a-40a M)E^3-
(5130+17088a-7920a M)E^2 \nonumber \\
\hspace{-1em}&+\hspace{-1em}&(14389+77376a+ 32256a^2-42032a M-50304
 a^2 M+13504 a^2 M^2)E- \nonumber \\
\hspace{-1em}&\hspace{-1em} &(9945+61248a+ 32256a^2-46800a M-124032 a^2 M+
43200a^2 M^2)
\label{c.4}
\end{eqnarray}
The energy eigenvalues are the roots of the polynomial equation such
that the Hill determinant \cite{kill} \cite{bo} $D(E)$ vanishes
$$ D(E)=
\pmatrix{1-E & 1 & 0 & 0 & 0 & 0 &..\cr
   8Ma      & 5-E & 1 &0 & 0 & 0 &..\cr
   0 & 48(M-1)a & 9-E &1 & 0&0  &..\cr
   0 & 0 & 120(M-2)a & 13-E & 1&0&..  \cr
   0 & 0 & 0 & 224(M-3)a & 17-E&1 &..\cr
   0 & 0 &0 &0&..&..&..\cr} 
$$
If the infinite matrix is truncated at order $N$, its determinant
$D_N (E)$ is solution of the recurrence relation
\begin{eqnarray}
D_{N+1} (E)=(4N+1-E)D_N (E)-8aN(2N-1)(M+1-N)D_{N-1}(E)
\label{c.6}
\end{eqnarray}
therefore $ D_N (E)=(-1)^N \, P_N (E, a, M)$. \\
Let us insert the formal expansion $E_0(a)=1+\sum_{n=1} ^\infty d_n a^n $, 
where $d_n$ are unknown variables, into the polynomial $ P_k (E, a, M)$
and expand in powers of the coupling $a$
\begin{eqnarray}
P_k (E=1+\sum_{n=1} ^\infty d_n a^n, a, M)=
\sum_{n=0} ^\infty c_p ^{(k)} [{d}, M] a^p
\label{c.7}
\end{eqnarray}
The coefficients $c_p ^{(k)} [{d}, M]$ depend on $d_r$ up to $r=p$
and vanish if $d_r$ are the perturbative coefficients of the expansion
of the ground energy eigenvalue, and $k>p$. This basic property of the
polynomials  $ P_k (E, a, M)$ allows to translate the eqs.(\ref{d.1})
into recursion relations for the coefficients $c_p ^{(k)}[{d}, M]$,
which allow the exact evaluation of $d_n(M)$ in an automated way.
The perturbative expansion of other, even and odd, energy eigenvalues
may also be performed with small changes \cite{kill} .\\
For instance one obtains for the ground 
state eigenvalue 
(to save space, I  only quote the term of order $a^{20}$ ):
\begin{eqnarray*}
 E_0(a)=1-2 M a-M(2M-3)a^2-M(4M^2-15M+12)a^3-M(40M^3-  \nonumber \\
264M^2+516M-297)a^4-
 M(28M^4-279M^3+948M^2-\frac{5229}{4}M+612)a^5- ...
\end{eqnarray*}
\begin{eqnarray}
..M \lbrack 14739534473769094441397440311/ 131072 &-&
  15258905673899527021780026567/ 32768 \,M +  \nonumber \\
  1784658246229507614806783961/2048 \,M^2 &-&
  507856798436368302225354717/512 \, M^3 + \nonumber \\
  6366289590321020410675505193/8192 \, M^4 &-& 
  114826346450120328738858675/256 \, M^5 +  \nonumber \\
  25467941727184707982391625/128 \, M^6 &-& 
  17860681260918574154980695/256 \, M^7 + \nonumber \\
  5045855866084769519143395/256 \, M^8 &-& 
  290745864268625747137893/64 \, M^9 + \nonumber \\
  13767162352069961178387/ 16 \, M^{10} &-& 
  2148046668981444052251/ 16 \,M^{11}  +  \nonumber \\
  275387290187286326553/ 16 \, M^{12} &-&
  1798144942712999139 \, M^{13} + \nonumber  \\
  150720427787130918 \, M^{14} &-& 
  9879984325256823 \, M^{15} + \nonumber  \\
  972416331565443/2 \, M^{16} &-& 
  16789635150822\,M^{17} + \nonumber  \\
  359298964716 \, M^{18} &-&  3534526380 \,M^{19} \rbrack \, a^{20}   +...
\label{c.8}
\end{eqnarray}
I checked the coefficients $d_n(M)$ up to $n=11$ by performing the
the regular perturbative expansion for $log[\psi(x)]$, a rather
efficient method for not very large order. Simple checks of the
coefficients $d_n(M)$ for higher orders are provided by evaluating
them for positive integer values of $M$, where they reproduce
the easily obtainable expansions of eqs.(\ref{a.16}).\\

\section{Concluding remarks.}
The properties of anharmonic oscillators perturbed by polynomial potentials
which correspond to quasi-exactly solvable models  are peculiar. The
perturbative expansions of the eigenvalues have a finite radius of 
convergence, which evades the general situation. In Sect.3, it was indicated
that properties of quasi-exactly solvable models may be useful to
the study of more general non-quasi-exactly solvable models. More
specifically, one may evaluate in exact, automated way, 
the perturbative expansions of energy eigenvalues. It would be very
interesting to know whether the radius of convergence of these expansions
collapses to zero, as soon as $M$ differs from a positive integer, or
there exist other real values of $M$ where such radius is finite. If this
were the case, quasi-exactly solvable models would have the additional
merit of suggesting ways of dealing with polynomial perturbations.
The answer to this question requires standard methods of analysis
of coefficients of the perturbative expansions which I hope to report
in a future work.\\
After the present letter was completed, I saw the recent paper by M.Znojil
\cite{zno} , which addresses similar issues, with different techniques and
an old letter by A.V.Turbiner and A.G.Ushveridze \cite{turbi} where
a subset of the investigation here reported in Sect.2 was performed.

\section{Acknowledgements}

I thank G.Burgio, M.P.Manara,P.L.Rigolli for useful discussions,
R.De Pietri for help with programming with Mathematica, P.Butera
for reading the manuscript and A.G.Ushveridze who started my interest
in quasi-exactly solvable models.

\end{document}